\documentclass[preprint,showpacs,preprintnumbers,amsmath,amssymb]{revtex4}
\usepackage{graphicx}
\usepackage{dcolumn}
\usepackage{bm}
\begin{document}
\preprint{}

\title{Kinetics of Spinodal Phase Separation in Unstable Thin Liquid Films}

\author{Rajesh Khanna$^{\ast}$, 
        Narendra Kumar Agnihotri$^{\ast}$,
        Manish Vashishtha}
 \email{rajesh@chemical.iitd.ac.in, agnik@gmail.com, mvche74@gmail.com}
 \affiliation{Department of Chemical Engineering, 
              Indian Institute of Technology Delhi, New Delhi -- 110016, 
              India.}
\author{Ashutosh Sharma}
 \email{ashutos@iitk.ac.in}
 \affiliation{Department of Chemical Engineering, 
              Indian Institute of Technology Kanpur, Kanpur -- 208016,  
              India.}
\author{Prabhat K. Jaiswal$^{\ddag}$ and Sanjay Puri}
 \email{prabhat.jnu@gmail.com, puri@mail.jnu.ac.in}
 \affiliation{School of Physical Sciences, 
              Jawaharlal Nehru University, New Delhi -- 110067, 
              India.}

\date{\today}

\begin{abstract}
We study universality in the kinetics of spinodal phase separation in
unstable thin liquid films, via simulations of the thin film equation.
It is shown that, in addition to morphology and free energy, the number 
density of local maxima in the film profile can also be used to identify 
the early, late and intermediate stages of spinodal phase separation. 
A universal curve between the number density of local maxima and
rescaled time describes the kinetics of the early stage in $d=2,3$.
The Lifshitz-Slyozov exponent of $-1/3$ describes the kinetics of 
the late stage in $d=2$ even in the absence of coexisting equilibrium
phases.
\end{abstract}


\maketitle
The kinetics of spontaneous phase separation, usually 
referred to as \emph{spinodal decomposition}, is of interest in 
far-from-equilibrium systems in diverse areas ranging from materials 
science and biological physics to cosmology and astrophysics  
\cite{pw09}. 
Spinodal phase separation is characterized by a combination of early, 
intermediate and late stages.
The early stage is marked by amplification/relaxation of initial 
fluctuations and dominance of the fastest amplifying
mode which emerges as the spinodal wave. 
Further amplification and saturation of the spinodal wave leads to 
emergence of new phases and begins the late stage of phase separation.
Domains of different phases grow as larger domains feed on 
smaller domains via diffusion or advection.
The intermediate stage features a mix of formation of new domains as well 
as growth of existing domains.
The dynamics of the late stage is well understood, and has been 
successfully explained by Lifshitz-Slyozov (LS) theory 
\cite{ls61}
and its variants. 
In contrast, the dynamics of the early and intermediate stages remains 
poorly understood. 
For example, it is not easy to study the time-dependence of the domain size
as given by correlation functions or structure factors, in the absence of 
well-defined phases.
Experimental access to these stages suffers from the added difficulty
of detecting low-amplitude  and multi-modal fluctuations. 
The present letter investigates the early and intermediate stages of 
spinodal phase separation in the technologically and scientifically
important system of supported unstable thin liquid films 
\cite{mat09}.
Our study is simulations of a model thin film equation
\cite{ruck74}.
We discuss the unique universal features of these stages and
present a useful marker to track their kinetics.
The LS growth scenario for the thin film system is also validated for 
the late stage. 
Interestingly, many different growth laws have been reported for 
the late stage
\cite{pfg03,wit09}.
The similarity between the thin film equation and the Cahn-Hilliard
(CH) equation
\cite{vsm93}
indicates that these unique features are likely to be shared by other
spinodally phase-separating systems.

\begin{figure}[htbp]
\begin{center}
\includegraphics*[width=0.8\textwidth]{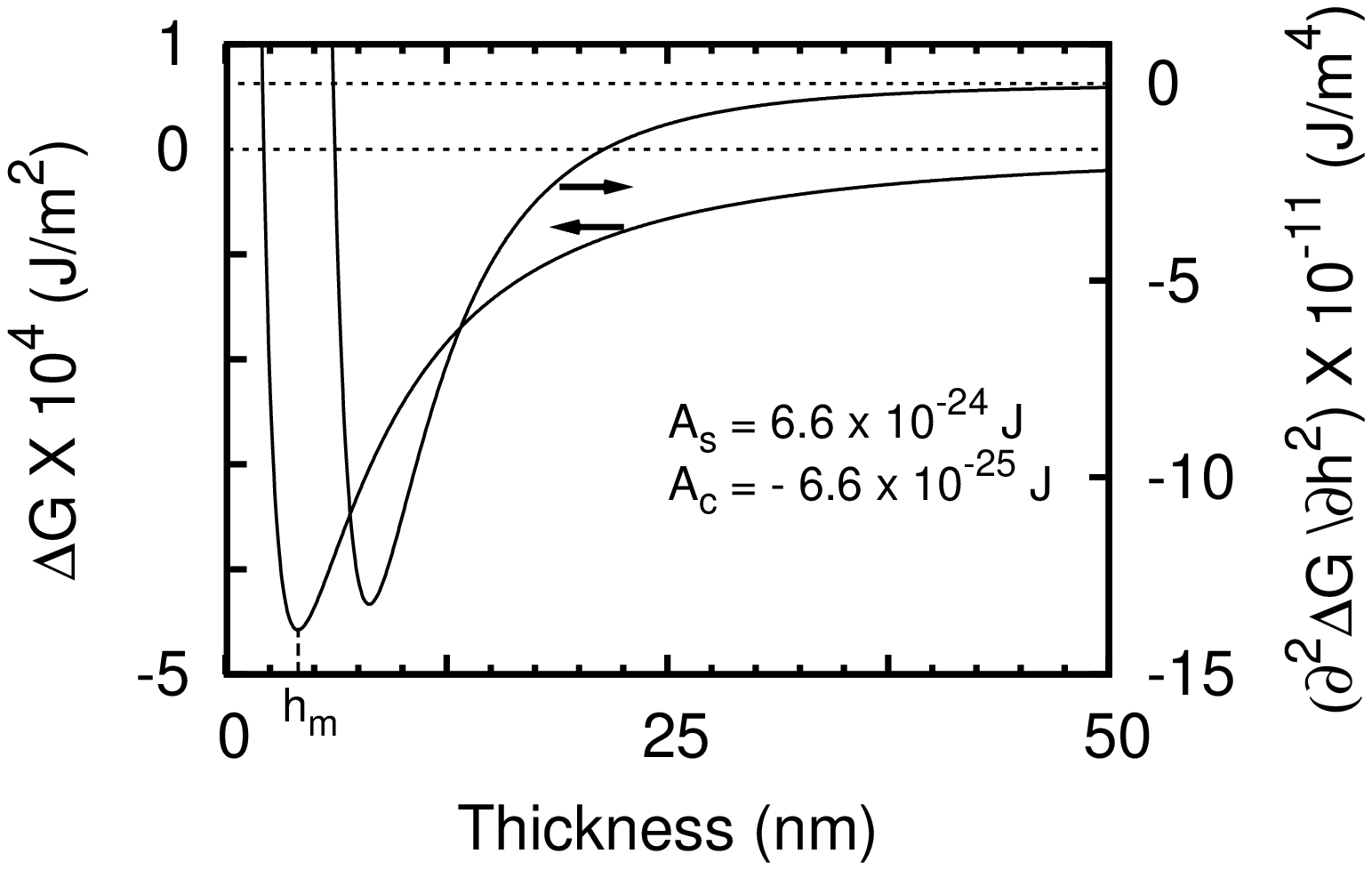}
\end{center}
\caption{Variation of the free energy (per unit area) of 
         a film ($\Delta G$) and force per 
         unit volume ($\partial^{2} \Delta G/ \partial h^{2}$) 
         with film thickness. 
         The parameter values are $R = -0.1$ and $\delta = 5 nm$.
         Spinodal phase separation occurs for $h_{0} > 6.25$ nm, where 
         $\partial^{2} \Delta G/ \partial h^{2} < 0$. 
         }
\label{fig:fig1}
\end{figure}

As in all spinodal processes, random fluctuations in the free surface of 
initially flat, supported thin liquid films ($< 100$ nm) grow and evolve 
into two distinct phases (viz., a low-curvature and thinner flat film
phase and a thicker high-curvature droplet phase) whenever $\Delta G$ 
shows a minimum and the spinodal parameter, 
$\partial^{2} \Delta G/ \partial h^{2}|_{h = {h_0}} < 0$ 
(Fig.~{\ref{fig:fig1}}). 
Here, $\Delta G$ is the excess intermolecular free energy (per unit area), 
$h$ is the film thickness, and $h_0$ is the average thickness 
\cite{as93, as98}.
Notice that the double-tangent construction for $\Delta G$ in 
Fig.~{\ref{fig:fig1}} shows that the film phase-separates into phases 
with $h = h_{m}$ and $h = \infty$. 
This should be contrasted with the usual phase-separation problems such
as segregation of binary mixtures which are described by a double-well 
potential, i.e., there are two possible values for the equilibrium 
composition
\cite{ab94}.
Actually, the droplet phase remains bounded due to overall volume
conservation and so does the maximum thickness.
One way to address this is to treat the droplet phase as a \emph{defect} 
with changing thickness rather than a \emph{true equilibrium phase}.

We study the thin film equation which models the spatio-temporal evolution 
of the film's surface in supported thin liquid films. 
This is derived by considering a thickness-dependent excess
intermolecular energy $\Delta G(h)$, and simplifying the equations 
of motion under the \emph{lubrication approximation} 
{\cite{ruck74}}. 
The resulting equation can be written as 
a $CH$ equation with a thickness-dependent effective mobility 
{\cite{vsm93}},
$M(h) = h^{3}/(3\mu)$ (corresponding to Stokes flow with no slip). 
The total free energy is 
$F_s[h] = \int [\Delta G(h) + \gamma (\vec{\nabla} h)^2/2] d\vec{x} \equiv F_e + F_i $
where, $F_{e}$ denotes the net excess free energy and $F_{i}$ denotes
the interfacial free energy.
In the above expressions, $\gamma$ and $\mu$ refer to surface tension and
viscosity of the liquid film, respectively. 
The corresponding CH equation is
\begin{equation}
\frac{\partial h}{\partial t}  
 = \nabla \cdot 
   \left[\frac{h^{3}}{3 \mu} \nabla
    \left( \frac{\partial \Delta G}{\partial h} - \gamma \nabla^{2}h
    \right)
   \right], 
\label{eq:dtfe}
\end{equation}
where all gradients are taken in the plane of the substrate.
Simulations were done for a variety of forms of $\Delta G(h)$
to uncover universal features of the early stage, if they exist. 
Here, a long-range van der Waals attraction due to the 
substrate, and a comparatively short-range van der Waals repulsion 
provided by a nano-coating on the substrate 
\cite{as96},
is  chosen to illustrate the  results. 
The corresponding $\Delta G$ is shown in Fig.\ {\ref{fig:fig1}}, 
and has the form 
$\Delta G = -A_{c}/12\pi h^{2} -A_{s}/12\pi(h+\delta)^{2}$. 
Here, $A_{s}$ and $A_{c}~(= R A_s)$ are the effective Hamaker constants 
for the system, which consists of the fluid bounding the film from the 
top, film fluid and a solid substrate ($s$) or coating material ($c$).
The thickness of the nano-coating is $\delta$. 
Mean film thicknesses falling in the spinodally unstable regime, given 
by $\partial^{2} \Delta G/ \partial h^{2}|_{h = {h_0}} < 0$, are also 
shown in Fig.\ {\ref{fig:fig1}}.

Equation~(\ref{eq:dtfe}) is solved in the 
following non-dimensional form to reduce parameters:
\begin{equation}
\frac{\partial H}{\partial T} = \nabla \cdot \left[H^{3} \nabla
\left( \frac{2 \pi h^{2}_{0}}{|A_{s}|} \frac{\partial \Delta G}{\partial H} - \nabla^{2}H
\right) \right]. 
\label{eq:tfe}
\end{equation}
Here, $H$ is the non-dimensional local film thickness scaled with the mean 
thickness $h_0$; the coordinates along the  
substrate are scaled with the characteristic length-scale for the 
van der Waals case 
$\left( 2 \pi \gamma/ \mid A_{s} \mid\right)^{1/2} h^{2}_0$;
and non-dimensional time $T$ is scaled with  
$\left( 12 \pi^{2} \mu \gamma h_0^{5}/ A_{s}^{2}\right)$. 
The excess energy term is completely non-dimensionalized as 
\begin{equation}
 \frac{2 \pi h^{2}_{0}}{\mid A_{s} \mid} 
  \frac{\partial \Delta G}{\partial H} = \frac{1}{3} 
  \left[ \frac{1-R}{\left(H + D\right)^{3}}+ \frac{R}{H^{3}}\right], 
\end{equation}
where $D = \delta/h_{0}$ is the non-dimensional coating thickness.
The linear stability analysis of Eq.\ (\ref{eq:tfe}) predicts a 
dominant spinodal wave of wavelength,  
$ L_{M} = 4 \pi/{\sqrt{-\frac{2\pi h_0^{2}}{\mid A_{s} \mid}\frac{\partial^{2}\Delta G}{\partial H^{2}}\Big|_{H = 1}}}$.

We numerically solve Eq.~(\ref{eq:tfe}) in $d=2,3$
starting with an initial small-amplitude ($\simeq 0.01$) 
random perturbation about the mean film thickness $H = 1$. 
In $d=2$, the system size is $nL_{M}$ ($n$ ranges from 16 to several 
thousands). 
Periodic boundary conditions are applied at the lateral ends. 
A 64-point grid per $L_{M}$ was found to be sufficient when central 
differencing in space with half-node interpolation was combined with 
{\it Gear's algorithm} for time-marching, which is especially suitable
for stiff equations. 
The parameters $D$ and $R$ were chosen so that the film is
spinodally unstable at $H = 1$. 
An increase in $D$ represents a corresponding decrease in the 
dimensional film thickness ($h_{0}$) for a fixed coating thickness 
($\delta$).

\begin{figure}[htbp]
\begin{center}
\includegraphics*[width=0.75\textwidth]{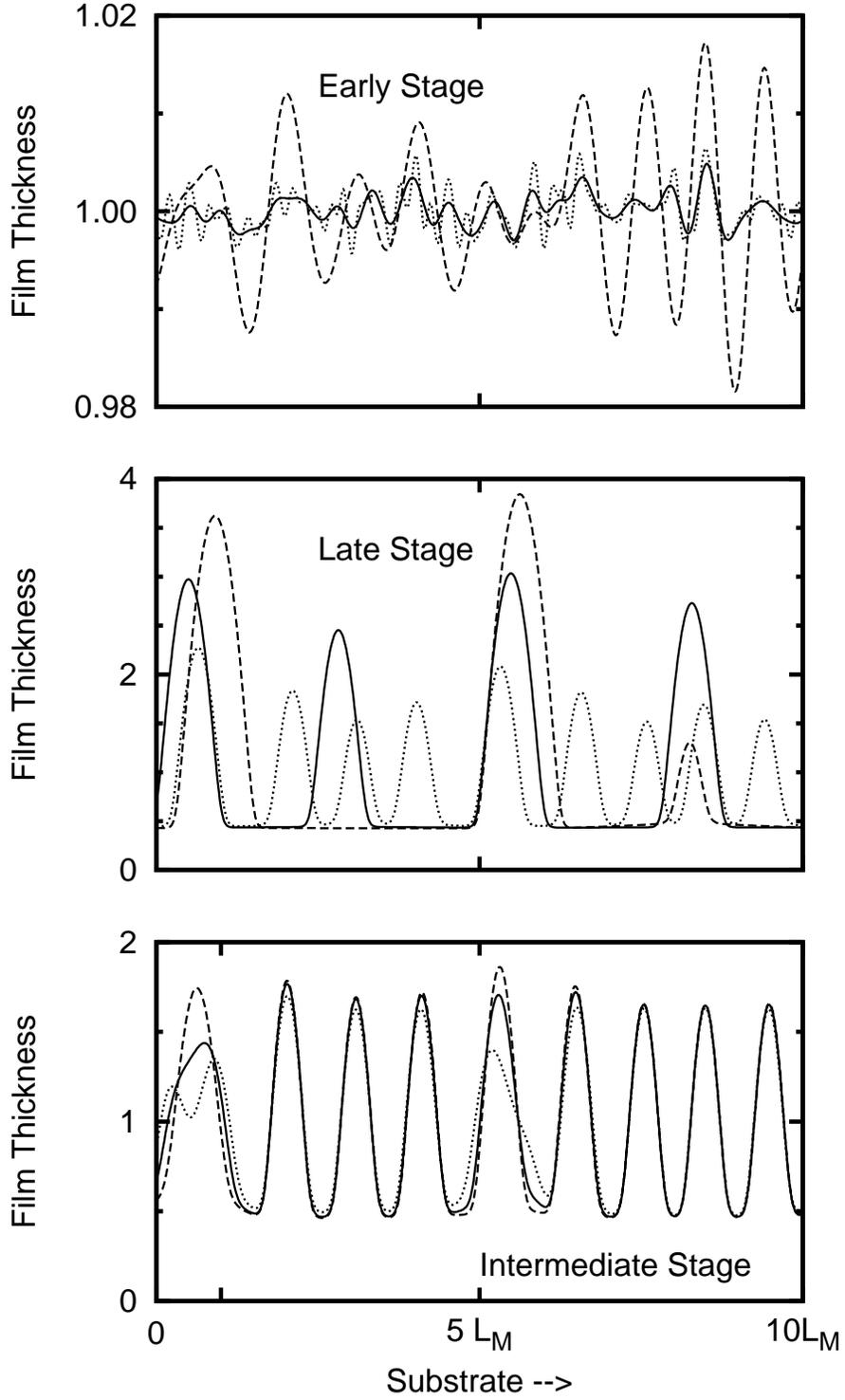}
\end{center}
\caption{Early, late and intermediate stages of phase
         separation in a supported thin film. 
         The dotted, solid and dashed lines refer to  
         non-dimensional times $0.6, 15$ and $760$ (top),
         $23000, 200000$ and $700000$ (middle),  
         $2310, 2550$ and $2725$ (bottom).  
         The parameters are $R = -0.1$ and $D = 0.5$.}
\label{fig:fig2}
\end{figure}

A morphology-based classification of phase separation is shown in 
Fig.~{\ref{fig:fig2}}. 
Curved-droplet defects and the flat-film phase can be clearly seen in the 
late stage (middle frame).
They are absent in the early stage (top frame), where the film surface is
characterized by a fluctuating wave of increasing wavelength, a result of 
relaxation of stable modes and growing dominance of the spinodal wave.
The intermediate stage (bottom frame) shows a mix of developing and 
fully-developed defects.
The effect of the spinodal wave can be clearly seen in the equi-spaced 
location and number of defects ($\sim 10$ defects in $10 L_{M}$).

\begin{figure}[htbp]
\begin{center}
\includegraphics*[width=0.8\textwidth]{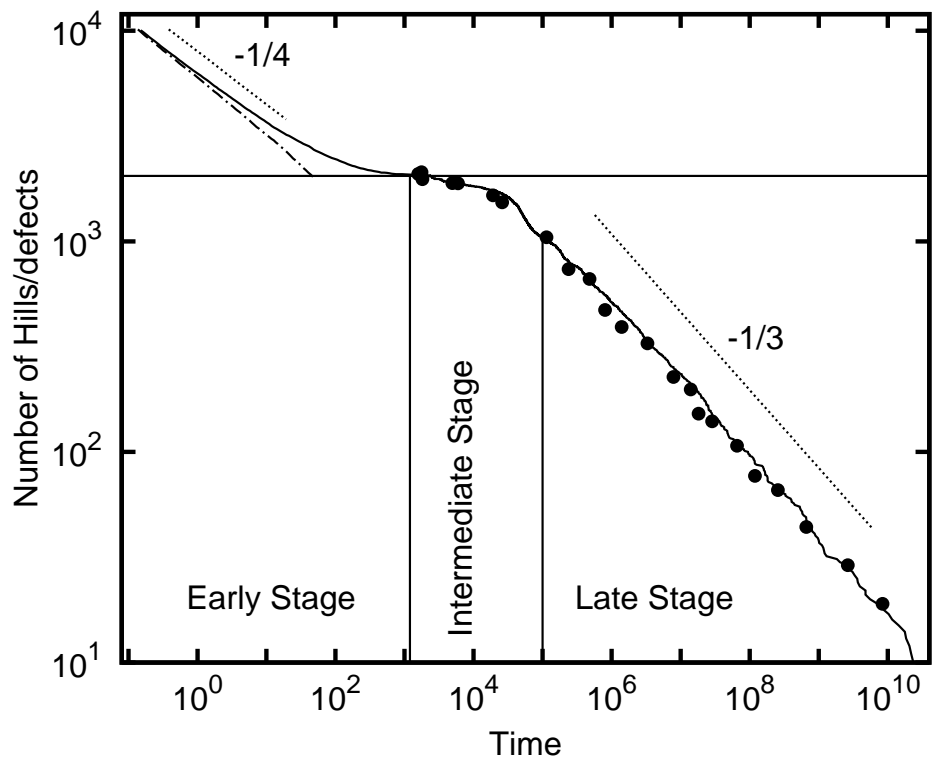}
\end{center}
\caption{Variation of the number of hills (defects) with non-dimensional 
         time for a system size of 2048 $L_M$.
         The filled dots show the corresponding results from Fourier 
         analysis.
         The dash-dot line represents the corresponding linear results.
         Dotted lines with  slopes of $-1/4$ and $-1/3$ show growth 
         exponents for the early and the late stages, respectively. 
 	 The parameters are $R = -0.1$ and $D = 0.5$.}
\label{fig:fig3}
\end{figure}

Figure~{\ref{fig:fig3}} shows the decrease in number of hills or
defects (local maxima in film profiles) with time. 
It shows three distinct stages:
an early stage with exponent $-1/4$, a late LS stage with exponent $-1/3$, 
and an intermediate stage. 
These stages coincide with those found by tracking the morphology and
validate the number of hills as a good marker of kinetics.
The early stage ends when the number becomes comparable to that for the 
spinodal wave, shown as a horizontal line in Fig.~{\ref{fig:fig3}}. 
The late stage is characterized by the start of $-1/3$ slope. 
The limits of the intermediate stage can then be defined accordingly.
An excellent matching of length-scales in the late stages as found by 
Fourier analysis (solid circles in Fig.~{\ref{fig:fig3}}) provides 
further support to the choice of number of hills as a marker.
As expected, Fourier analysis fails to provide any characteristic
length-scale in the early stages.
Experimental results for all stages can be consistently described by 
pooling early-stage data based on number of hills, and late-stage data 
based on Fourier analysis. 
Interestingly, linear theory 
(plotted as dash-dots in Fig.~{\ref{fig:fig3}}) only explains 
the beginning of the early stage. 
So, while the spinodal wave itself is given by linear theory, 
the kinetics leading to its emergence can only be described by nonlinear 
analysis. 

\begin{figure}[htbp]
\begin{center}
\includegraphics*[width=0.8\textwidth]{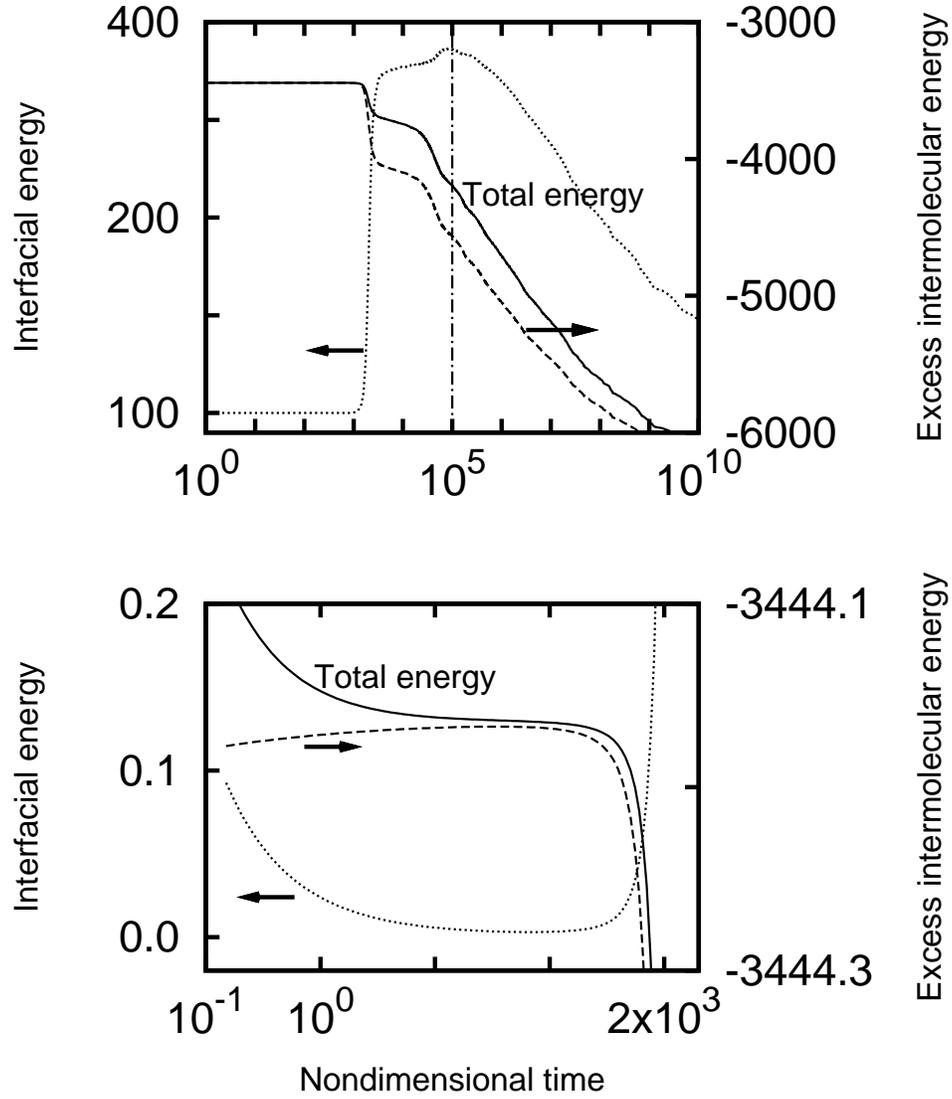}
\end{center}
\caption{Variation of the interfacial free energy ($F_{i}$), excess free 
         energy ($F_{e}$) and total free energy ($F_{s}$) with 
         non-dimensional time.
         The bottom frame shows the magnified view in the early stage.
 	 The parameters are $R = -0.1$ and $D = 0.5$.}
\label{fig:fig4}
\end{figure}

\begin{figure}[htbp]
\begin{center}
\includegraphics*[width=0.8\textwidth]{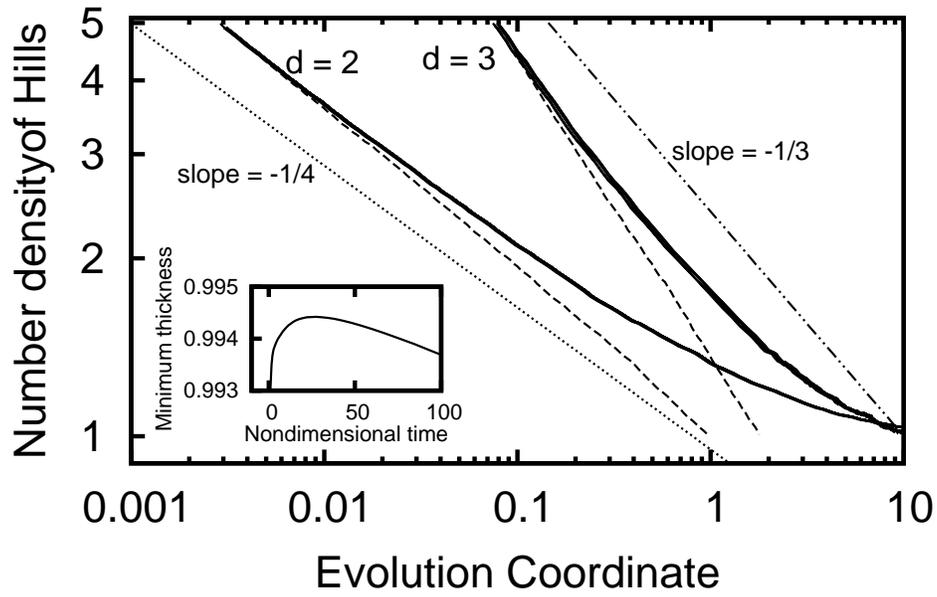}
\end{center}
\caption{Master curves in $d=2,3$ for variation of number density of 
         hills with the evolution coordinate.
         We superpose data for a wide range of potential and
         parameters, as described in the text. 
         The inset shows the variation of minimum thickness with $T$
         which is used in defining the evolution coordinate.}
\label{fig:fig5}
\end{figure}

These stages can be readily identified in the evolution of the free energy 
also (Fig.\ {\ref{fig:fig4}}). 
In the early stage, there is not much change in $F_s$, $F_e$ and $F_i$. 
This stage ends with a sharp increase in $F_i$ and a corresponding 
decrease in $F_e$.
In the late stage, there is a smooth decrease in both $F_i$ and $F_e$. 
$F_{i}$ depends on the sum of squares of local slopes.
The random perturbations at the start result in a high value of the
interfacial energy. 
The smoothening of these leads to a decrease in local slopes and 
$F_{i}$ (bottom frame of Fig.~{\ref{fig:fig4}}). 
The subsequent emergence of the high-curvature defects and their growth 
lead to increase in $F_{i}$.
On the other hand, $F_{e}$ depends on the local thickness and decreases 
with reduction of thickness for films in the spinodal regime. 
The initial increase in the minimum thickness 
(inset of Fig.~{\ref{fig:fig5}}), a result of reducing 
amplitude of stable components of the surface perturbation, 
leads to a slight increase in $F_{e}$.
As the minimum thickness passes through a maximum and starts
decreasing (inset of Fig.~{\ref{fig:fig5}}), 
$F_{e}$ also starts decreasing 
(bottom frame of Fig.~{\ref{fig:fig4}}). 
The total free energy decreases all the time, showing the spontaneous
nature of the phase separation.
Again, an analysis of the film profiles shows that the
classification based on free energy is consistent with the
analysis based on the number of hills as well as morphology.
We remark that $F_{i}$ is the best marker for defining the
intermediate stage. 
The number of hills do not provide such a clear indication of the 
start and end of the intermediate stage.
A morphological analysis can demarcate the intermediate
stage well but only if it is exhaustive and covers the complete profile. 

We have also obtained results for the early stage for several other 
parameter values ($R$ and $D$), system sizes ($nL_{M}$) and other force 
fields, including an uncoated substrate ($D = 0$) and a combination of 
van der Waals and polar interactions
\cite{as96}.
These also show the same feature of steady decrease in number 
of hills (with exponent $\sim -1/4$, similar to Fig.~{\ref{fig:fig3}}), 
with results separated only in time scales (results not shown). 
The effect of system size is easily removed by plotting the number density 
of hills (number of hills per $L_M$) vs. $T$ rather than number of hills. 
One finds that the kinetics is independent of the system size 
(results not shown).
This independence is especially useful for experimentalists who
need not worry about the lateral size of the sample.

A master curve to describe all these results will strongly indicate a 
universal early-stage kinetics.
How does one rescale the results to arrive at the master curve?
It is known that initial fluctuations greatly influence the 
kinetics for film  thicknesses at the edge of the spinodal regime or 
in the {\it defect sensitive spinodal region} (DSSR), as opposed to 
thicknesses in the {\it deep inside spinodal region} (DISR).
\cite{as07}. 
A new evolution coordinate found by dividing $T$ by a time,
$T_{i}$, at which different films are at a given morphological event 
during their evolution, will account for the different initial fluctuations.
There can be many such choices of events, with the emergence of the 
flat-film phase being the most obvious one. 
However, the increasing dominance of the repulsive force field slows down 
the dynamics near this event, leading to uncertainty in estimating the 
corresponding times.
Another event, which provides a much sharper estimate, is found by 
tracking the minimum thickness of the evolving film: 
$T_{i}$ can be chosen as the time at which it reaches its largest value 
(see inset of Fig.\ {\ref{fig:fig5}}).
Indeed all results, when plotted using this evolution coordinate, 
collapse on a master curve as shown in Fig.\ {\ref{fig:fig5}}. 
This shows that the early stages of phase separation in thin liquid films, 
as marked by the number density of hills, can be represented by a 
universal curve with exponent $-1/4$. 
Further confirmation of this universality is provided by results for the 
$d=3$ version of Eq.\ (\ref{eq:tfe}). 
They also show the same universality but with an exponent closer 
to $-1/3$. 
This is consistent with results from the phase separation of mixtures 
\cite{pw09}.
The early formation and coalescence of domains is known to have a 
dependence on dimensionality.

In summary, we conclude that the number density of local maxima
(defects) and not the usual measures like structure factors and
correlation functions can be used to identify the early, late and 
intermediate stages of spinodal phase separation in thin films. 
The kinetics of the early stages for a wide range of potentials and
parameters can be described by a universal 
curve between the number density of defects and a rescaled time in $d=2,3$.
This process has strong similarities with the usual phase-separation
processes except for the absence of coexisting phases.
Still, LS exponent of $-1/3$ governs the late stages of domain growth.
Our results in this letter will facilitate the understanding of many 
experimental results for phase separation in thin films.
We also hope that our predictions of universality in early-stage dynamics 
will be put to experimental test.

R.\ K. acknowledges the support of Department of Science and
Technology, India.

\bibliography{draft}

\newpage

\newpage

\newpage

\newpage

\newpage

\end{document}